\documentclass[reprint, amsmath,amssymb, aps,prl,
superscriptaddress,nofootinbib]{revtex4-2}
\usepackage{graphicx}
\usepackage{dcolumn}
\usepackage{bm}
\usepackage{amssymb}
\usepackage{amsmath}
\usepackage{appendix}
\usepackage{xcolor}
\usepackage[mathscr]{euscript}
\usepackage{mathtools}
\usepackage{todonotes} 
\usepackage{hyperref}
\usepackage{xcolor}
\hypersetup{
    colorlinks,
    linkcolor={blue!50!blue},
    citecolor={blue!50!blue},
    urlcolor={blue!80!black}
}
\DeclarePairedDelimiter\bra{\langle}{\rvert}
\DeclarePairedDelimiter\ket{\lvert}{\rangle}
\DeclarePairedDelimiterX\braket[2]{\langle}{\rangle}{#1 \delimsize\vert #2}

\begin{document}

\title{Topological Bogoliubov quasiparticles from Bose-Einstein condensate in a flat band system}

\author{Zahra Jalali-Mola}
 \affiliation{ICFO - Institut de Ciencies Fotoniques, The Barcelona Institute
of Science and Technology, 08860 Castelldefels, Barcelona, Spain}

\author{Tobias Grass}
\affiliation{ICFO - Institut de Ciencies Fotoniques, The Barcelona Institute of Science and Technology, 08860 Castelldefels, Barcelona, Spain}%
\affiliation{DIPC - Donostia International Physics Center, Paseo Manuel de Lardiz{\'a}bal 4, 20018 San Sebasti{\'a}n, Spain}
\affiliation{Ikerbasque - Basque Foundation for Science, Maria Diaz de Haro 3, 48013 Bilbao, Spain}

\author{Valentin Kasper}
\affiliation{ICFO - Institut de Ciencies Fotoniques, The Barcelona Institute of Science and Technology, 08860 Castelldefels, Barcelona, Spain}
\affiliation{Nord Quantique, 3000 boulevard de l'Université (P1-ACET), Sherbrooke J1K 0A5, QC, Canada}

\author{Maciej Lewenstein}
\affiliation{ICFO - Institut de Ciencies Fotoniques, The Barcelona Institute of Science and Technology, 08860 Castelldefels, Barcelona, Spain}
\affiliation{ICREA, Pg. Llu\'is Companys 23, 08010 Barcelona, Spain}

\author{Utso Bhattacharya}
\affiliation{ICFO - Institut de Ciencies Fotoniques, The Barcelona Institute of Science and Technology, 08860 Castelldefels, Barcelona, Spain}

\date{\today}

\begin{abstract}
For bosons with flat energy dispersion, condensation can occur in different symmetry sectors. Here, we consider bosons in a Kagome lattice with $\pi$-flux hopping, which in the presence of mean-field interactions exhibit degenerate condensates in the $\Gamma$- and the $K$-point. We analyze the excitation above both condensates and find strikingly different properties: For the $K$-point condensate, the Bogoliubov-de Gennes (BdG) Hamiltonian has broken particle-hole symmetry (PHS) and exhibits a topologically trivial quasiparticle band structure. However, band flatness plays a key role in breaking the time reversal symmetry (TRS) of the BdG Hamiltonian for a $\Gamma$-point condensate. Consequently, its quasiparticle band structure exhibits non-trivial topology, characterized by non-zero Chern numbers and the presence of edge states. Although quantum fluctuations energetically favor the $K$-point condensate, the interesting properties of the $\Gamma$-point condensate become relevant for anisotropic hopping. The topological properties of the $\Gamma$-point condensate get even richer in the presence of extended Bose-Hubbard interactions. We find a topological phase transition into a topological condensate characterized by high Chern number and also comment on the realization and detection of such excitations.
\end{abstract}
\maketitle

\emph{Introduction.}
The discovery of topological band structures has led to an entire new field of physics on topological properties of quantum matter \cite{Hasan2010,Qi2011}. The non-trivial topology of bulk Bloch bands in topological insulators and superconductors possess gapless edge states which are robust against local impurities and give rise to responses that are precisely quantized. A prototypical example of a topological insulator is a fermionic two-dimensional integer quantum Hall system \cite{von86}, which exhibits a quantized
Hall conductivity proportional to the non-zero Chern number of the occupied bands. The non-zero value of the Chern number originates from the breaking of TRS due to an applied magnetic field and is responsible for the existence of uni-directional gapless chiral modes  propagating along the edges of the system, however the bulk is completely insulating. 
In the presence of interactions, such systems may develop topological order, with even more striking phenomena such as anyonic quasiparticles \cite{Nayak2008}, possibly even for mean-field interactions as in the case of Majorana modes in a $p$-wave superconductor. 

However, the notion of topological protection is not tied  to fermionic systems only as interactions also open an avenue for probing topological band structure in bosonic systems, as observed in beautiful quantum gas experiments with cold bosonic atoms \cite{Aidelsburger2015}. While in such a case the non-trivial topology is already present on the level of the single-particle band structure and interactions are only a tool to fill the topological band with bosonic particles, there are other scenarios in which the topologically non-trivial behavior is induced only by the interactions. In particular, bosonic condensates with broken TRS can give rise to collective excitations which exhibit topological bands \cite{DiLiberto2016}. This exotic phenomenon may happen in degenerate bands and has recently been observed by preparing a Bose-Einstein condensate (BEC) within the $p$-band of a honeycomb lattice \cite{Wang2021}. 

An extreme case of band degeneracy is a flat band where many single-particle states are dispersionless and localized. 
Quantum systems with a flat energy dispersion have recently attracted a lot of attention, especially due to the realization of flat bands in magic-angle twisted bilayer graphene \cite{Cao2018}, as well as in  synthetic systems \cite{Leykam_rev_flat, Zhai_PRL2012,Ueda_2015,Toram_PRB2021,Torma_PRL2021}.
Primary questions addressed in these works concern how due to a vanishing of kinetic energy,
the transport properties of a flat band are determined by
the quantum geometry of Bloch states and out of many degenerate states which one favors a stable BEC. A good intuition for the behavior of a BEC can usually be obtained from a mean-field treatment, but the flat band scenario comes with some caveats: The huge single-particle degeneracy of the band may survive on the mean-field level, and may only be resolved by the contribution of fluctuations through a mechanism known as order by disorder \cite{Villain1980,Barnett2012}. For instance, a mean-field Bose-Einstein condensate can select the $\Gamma$-point or the $K$-point, as well as an extensive number of configurations with broken translational symmetry, but the degeneracy is lifted through quantum and/or thermal fluctuations, cf. Ref.~\cite{Zhai_PRL2012}.

\begin{figure}[t]
	\centering
	\includegraphics[width = 1\linewidth]{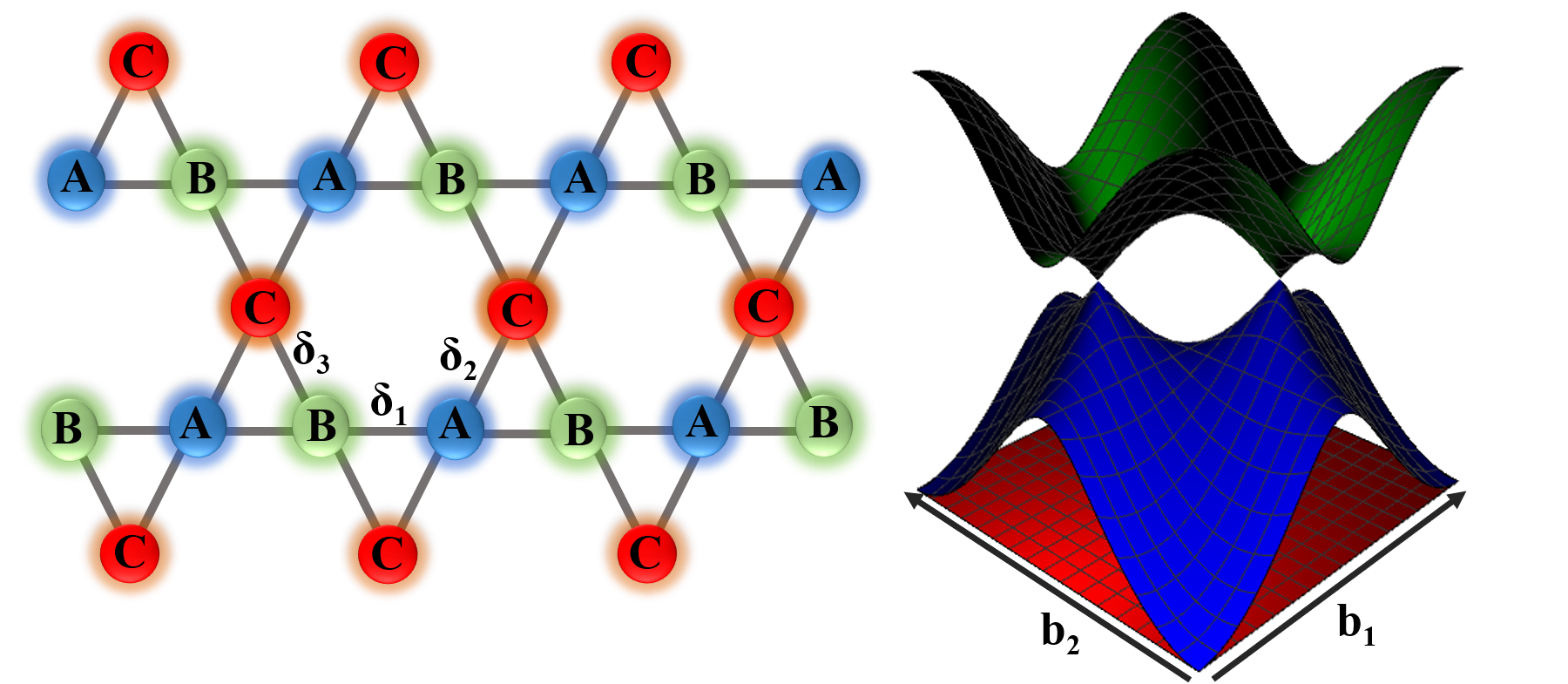}
	\caption{Kagome lattice with three different sublattices in the left panel, and single-particle dispersion relation of the tight-binding Hamiltonian in Eq.~\eqref{TB_K.eqn} for $t>0$ in the right panel. Here, $\delta_i$ represents the distance between two different nearest neighbor sublattices, and $b_i$ stands for the reciprocal lattice vectors in momentum space. The vertical axis in the right panel is the dimensionless quantity $E_0(k)/t$.}
	\label{fig_h0}
\end{figure}

Due to the suppression of kinetic energy, topological properties are expected to play a key role in determining the nature of condensates in flat band systems. For the investigation of topological properties of flat bands, the distinction into two categories of flat bands turns out to be useful: singular and non-singular~\cite{Leykam_rev_flat, Rhim_Flat_2019} bands, depending on the compact localized properties of the Bloch functions. In a singular flat band, removing any degeneracy or band crossing contaminates the flatness, and the flat bands can acquire non-zero Chern numbers~\cite{Rhim_Flat_2019}. In contrast, in non-singular flat bands, the zero width of the band is a robust property. Singular flat bands are supported, for instance, by lattices with Kagome geometry, as shown in Fig.~\ref{fig_h0}. Optical lattices with Kagome geometries~\cite{opt_Kagome_Vishwanath,Maciej_science2011} and their topological properties~\cite{Kagome_topo_2012,Kagome_Chisnell_2015,Li2020_kagome_topo,Kagome_topo_2022} have come under scrutiny in recent years. 
Condensation in the flat band is possible via artificial gauge fields making it the energetically lowest band. Schemes to produce such gauge fields have been developed for a variety of synthetic quantum systems \cite{Lin2009,Maciej_rev,RMP_gauge,Maciej_book,Lu2014_review,Goldman2016,RMP_Ozawa,RevMod_topological}.

In this letter, we consider the scenario of a Kagome lattice with a synthetic $\pi$-flux, that is, with a real but positive hopping amplitude. As shown in Fig.~\ref{fig_h0}, the lowest band of this system is the flat one. We then see that out of this degenerate manifold, the presence of onsite and nearest neighbor interactions selects two possible translationally invariant mean-field condensates, at the $\Gamma$-point and at the $K$-point. Then the collective excitations above these condensates on the level of a quadratic BdG Hamiltonian are studied. We summarize the two key findings from our analysis at the outset: (a) We show that the symmetry properties of the BdG Hamiltonian depend crucially on the wave vector of the condensate. Strikingly, although the single-particle bands are topologically trivial, it is band flatness and the onsite interactions that conspire to render the condensate of the $\Gamma$-point as non-trivial, which may further undergo a topological phase transition into bands with higher Chern number through the presence of nearest neighbor interactions~\cite{Kagome_NN_Exp.}. (b) While the non-trivial condensate is unstable with respect to the zero-point fluctuations in the case of a Kagome lattice with isotropic hopping, we explicitly argue how one can overcome this challenge by engineering  anisotropic hopping terms which can enforce the topological condensate ~\cite{Moessner2018}, with possible scope for experimental realization with ultracold artificial Kagome lattice.
   
\emph{System.}
The Kagome structure is composed of three sublattices $A$, $B$, and $C$ (see the left panel of Fig.~\ref{fig_h0}), and the tight-binding Hamiltonian of particles on such a lattice is $H_0=t \sum_{\substack{\langle i,j \rangle \\ \alpha,\beta}} d^{\dagger}_{i,\alpha} d_{j,\beta}-\mu \sum_{i,\alpha} d^{\dagger}_{i,\alpha} d_{i,\alpha}$, where $d_{i,\alpha}$~($d^\dagger_{i,\alpha}$) are the annihilation~(creation) operator for particles on sublattice $\alpha$ located at position $i$. The amplitude of hopping between nearest neighbors $ \langle i,j \rangle$ is $t>0$, and the $\pi$-flux is accounted by the sign of the hopping term.
In momentum space, the Hamiltonian reads
\begin{equation}
    \label{TB_K.eqn}
    H_0(k) =\begin{bmatrix}
    -\mu &2t \cos k_1& 2t\cos k_2 \\
    2t\cos k_1& -\mu & 2t\cos k_3 \\
    2t\cos k_2& 2t\cos k_3& -\mu
    \end{bmatrix},
\end{equation}
with $k_i=\textbf{k} \cdot \boldsymbol{\delta}_i$ where $\boldsymbol{\delta}_i$ is the vector between two nearest neighbors, as defined in the left panel of Fig.~\ref{fig_h0}. In the rhomboidal Brillouin zone (right panel of  Fig.~\ref{fig_h0}), we observe a band touching at $\Gamma=(0,0)$ between the flat band (red) and the middle band (blue), and between the middle band and the upper band (green) at the $K$ and $K'$ points. 
For the interactions, we consider repulsive on-site interactions $U$, and also account for possible nearest neighbor interactions $V$ in the interaction Hamiltonian given by $H_I=\frac{U}{2} \sum_{i, \alpha} \left( n_{i,\alpha} n_{i,\alpha}-1\right)+\frac{V}{2}\sum_{\substack{\langle i,j \rangle \\ \alpha,\beta}} n_{i,\alpha} n_{j,\beta}$, where the density operator is $n_{i,\alpha}= d^{\dagger}_{i,\alpha} d_{i,\alpha}$.

In a mean-field treatment, the bosonic operators are replaced by their expectation values, $\langle d_k\rangle=d^\circ_{k}=(\psi_{k,A}, e^{i\phi_{k,B}}\psi_{k,B},  e^{i\phi_{k,C}}\psi_{k,C})$, and the energy becomes a function of the real-valued mean-fields $\psi_{k,\alpha}$ and $\phi_{k,\alpha}$, which are found by minimizing the energy. We concentrate on translationally invariant mean-field solutions in which condensation occurs in a single mode, denoted $k_{cp}$. We use the hopping parameter $t$ as a unit of energy, and the chemical potential $\mu$ serves to adjust the condensate density $\rho=\sum_\alpha |\Psi_{k_{cp},\alpha}|^2$, which we set to 1. The interaction parameters $U,V$ are considered tunable, and we find that qualitatively two regimes must be distinguished:

(i) $U>2V$: In this, physically easily realizable scenario, the mean-field energy,   $E_{Min}=-2t\rho+(U+4V)\rho^2/3$, occurs in two modes, at the $\Gamma$ and $K$ points. In both modes, the mean-field solution is uniform in the sublattices, i.e. $|\psi_\alpha|^2=\rho/3$. An important difference between the two degenerate condensates is the complex phase in the $\Gamma$-point condensate, in contrast to the real-valued amplitude of the condensate in the $K$-point. Specifically, the two solutions read:
 \begin{align}
     \label{con_g_nonrot_1.eqn}
      & d^\circ_\Gamma=\frac{1}{\sqrt{3}}(1,-e^{i\pi/3}, e^{2i \pi/3}),\\
      \label{con_K_nonrot.eqn}
      & d^\circ_K=\frac{1}{\sqrt{3}}(1, 1, -1).
\end{align}

(ii) $U<2V$: In this case, the energy is minimal only at the $\Gamma$-point. As a result of strong nearest neighbor interaction, the condensate is not uniform in the sublattice anymore. In addition, we mention that at $U=2V$ the infinite degeneracy of the flat band, which is typically removed by the interactions, reappears as a consequence of the competition between $U$ and $V$. However further investigations in this direction are beyond the scope of this Letter.

\emph{Bogoliubov quasiparticles.}
Despite the possibly very interesting physics which may occur for dominant $V$, in the following, we focus on the  more physical scenario (i).
To account for quantum fluctuations around the mean-field and obtain the excitations of the mean-field system, we split the operators into mean-field part and fluctuations, $d_k=d_{k_{cp}}^\circ+\delta d_k$, where $\delta d_k=(1-\delta_{k,k_{cp}})d_k$ is zero at the condensation point. To obtain a quadratic BdG Hamiltonian, $H_B(k)=\frac{1}{2}\sum_{\alpha,\beta}\Psi^\dagger_{k,\alpha} H^{MF}_{\alpha,\beta} \Psi_{k,\beta} + {\rm const.}$, we keep fluctuating terms only to second order and define Nambu spinors ${\bf \Psi}_k = (\Psi_{k_+},\Psi^\dagger_{k_-})^T$, in which the two components represent particle-like and hole-like part of the wave function, with $\Psi_{k_\pm} = (\delta d_{k_\pm,A},\delta d_{k_\pm,B},\delta d_{k_\pm,C})^T$ and $k_\pm=k_{cp}\pm k$. The kernel of the BdG Hamiltonian reads
\begin{align}
H_B=
   \label{MFH.eqn}
   \begin{bmatrix}
        H_0(k_{cp}+k)+\mathcal{H}_{0}(k) & H_{\Delta }(k) \\
       H^*_{\Delta }(k)  & H^T_0(k_{cp}-k)+\mathcal{H}^{ *}_{0}(k)
    \end{bmatrix},
\end{align}
where the diagonal part contains the tight-binding Hamiltonian $H_0(k)$ from Eq.~\ref{TB_K.eqn}
 and a mean-field contribution ${\cal H}_0(k)$ from the interactions. The off-diagonal terms $H_\Delta$ stem exclusively from the mean-field decomposition of interactions, see supplemental material~\ref{AppA} for explicit expressions. 

Diagonalization of the BdG Hamiltonian needs to account for the commutation relation of the Nambu spinors~\cite{Xiao_diag}, $[\Psi_k,\Psi^\dagger_{k'}]=\sigma_3 \delta_{k,k'}$, where $\sigma_3=\sigma_z \otimes I_3$ acts on Nambu space as the Pauli matrix $\sigma_z$.
The eigenmodes of the BdG Hamiltonian are obtained from the pseudo-Hermitian Hamiltonian $\sigma_3 H_B(k)$, and the transformation matrix $W(k)$ which diagonalizes $\sigma_3 H_B(k)$ satisfies the following relations:
 \begin{align}
    \label{Wpm.eqn}
    & W^{\dagger}(k) \sigma_3 W(k)=\sigma_3,\\
    \label{WHW.eqn}
    & W^\dagger(k) H^{MF}(k) W(k)= {\rm diag}[\boldsymbol\omega(k_+),\boldsymbol\omega(k_-)],\\
    \label{WzHW.eqn}
    & W^{-1} \sigma_3~H^{MF}(k) W(k)= \sigma_3~ {\rm diag}[\boldsymbol\omega(k_+),...,\boldsymbol\omega(k_-)].
\end{align}
Here, $\boldsymbol\omega(k)=[\omega_1(k),\omega_2(k),\omega_3(k)]^T$ represents lowest to highest eigenenergies at momentum $k$, respectively, for the Bogoliubov quasiparticles ($k_+$) and quasiholes ($k_-$).
The lowest energy band $\omega_1$ should have zero energy at the condensation point  which fixes the chemical potential $\mu$~\cite{SHI19981,Toram_PRB2021}.

In the BdG Hamiltonian, PHS and TRS are defined as ~\cite{Kawaguchi_2020_symmetry,Murakami_symmetry}
\begin{align}
    & \label{TR.eqn}
    \text{TRS:}~~~ H^*_B(k)=H_B(-k), \\
    \label{PH.eqn}
    & \text{PHS:}~~~ \sigma_1 H^*_B(k) \sigma_1=H_B(-k).  
\end{align}
We find that the symmetry properties of the BdG Hamiltonian depend on the choice of mean-field momentum: For $k_{cp}=\Gamma$, the complex-valued condensation parameters break TRS, while the diagonal blocks in the Bogoliubov Hamiltonian are the same, i.e. PHS is preserved. For $k_{cp}=K$, real-valued condensation parameters keep TRS intact, but the finite momentum of the condensate breaks PHS. 

\begin{figure}[t]
	\centering
	\includegraphics[width = 1\linewidth]{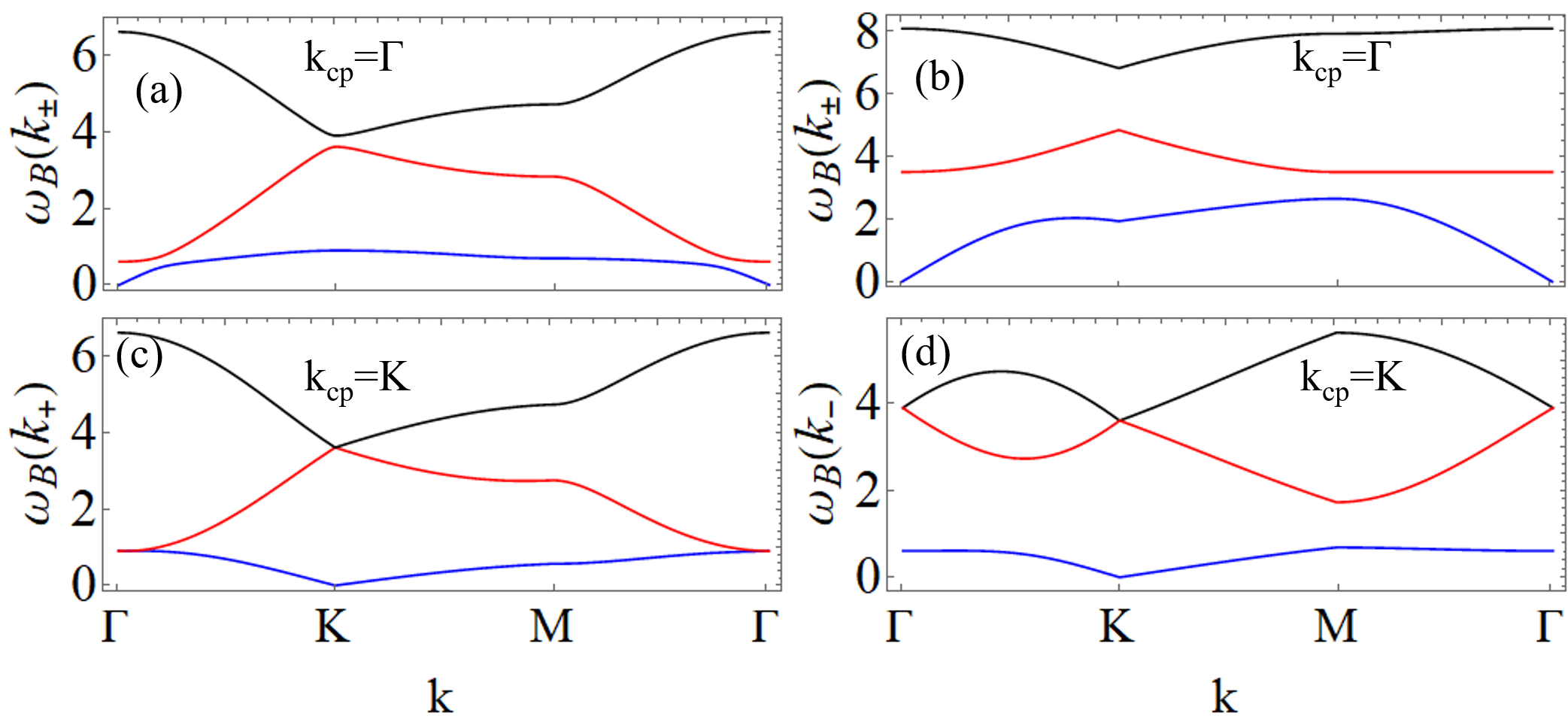}
	\caption{ Bogoliubov dispersions of quasiparticle and quasiholes, for different values of nearest neighbor interaction $V$ and fixed on-site interaction $U/t=3$. The value of nearest neighbor interaction  is $V/t=0.5$ in panels (a,c,d) and $V/t=4$ in panel (b). In panel~(a) and~(b), condensation occurs at $k_{cp}=\Gamma$. In this case, particle-hole symmetry leads to $\omega(k_+)=\omega(k_-)$. In panels~(c) and~(d), condensation occurs at $k_{cp}=K$. In this case, particle-hole symmetry is broken and we depict the dispersion of quasiparticles and quasiholes separately. In all panels, we have chosen $t=\rho=1$.}
	\label{fig_Bg_pm_mu}
\end{figure}

These symmetry properties have important consequences for the excitations whose spectra are plotted in Fig.~\ref{fig_Bg_pm_mu} for the two different $k_{cp}$. For $k_{cp}=\Gamma$, PHS makes particle, and hole spectra indistinguishable. They are plotted in  Fig.~\ref{fig_Bg_pm_mu}(a) and Fig.~\ref{fig_Bg_pm_mu}(b) for two different values of $V$. For $k_{cp}=K$, particle and hole spectra are different, and plotted separately in Fig.~\ref{fig_Bg_pm_mu}(c) and Fig.~\ref{fig_Bg_pm_mu}(d), for the same choice of $V$ as used in Fig.~\ref{fig_Bg_pm_mu}(a). The effect of TRS breaking is seen in Fig.~\ref{fig_Bg_pm_mu}(a) by the gap openings between all the bands in whole Brillouin zone, absent in Fig.~\ref{fig_Bg_pm_mu}(b), (c) and (d).

Broken TRS is also expected to have consequences for the topological properties of the collective modes, cf. Refs.~\cite{Ueda_2015,Brandes_PRA2015,PRA_2021_toint,Shindou_PRB2013_1,Shindou_PRB2013_2,wang2017dirac}. Band curvature and Chern numbers are defined as
\begin{align}
    \label{Berry.eqn}
    & B_m(k)=i\sum_{i.j}\epsilon_{ij}\bra{\partial_i W(k)}\sigma_3\ket{\partial_j W(k)}_{mm} (\sigma_3)_{mm},\\
    \label{Chern.eqn}
   & C_m=-\frac{1}{2\pi}\int_{BZ} d^2\boldsymbol{k} B_m(k). 
\end{align}
Here, $m$ stands for the Bogoliubov mode band index. We use the Fukui-Hatsugai-Suzuki method~\cite{Fukui_method} to evaluate the Chern number for Bogoliubov excitation bands, making use of  Eq.~\eqref{Chern.eqn}, and keeping in mind that the Chern number is not well-defined in the lowest band $\omega_1$ at the condensation point  $k=k_{cp}$, and applying a rotation $R(k)=diag \left(e^{i k_2}, e^{-ik_3},1,e^{i k_2}, e^{-ik_3},1\right)$ to make the BdG Hamiltonian periodic in the first Brillouin zone.
As expected, all Chern numbers are trivially zero for the condensate in $K$, but non-zero in the case of condensate in $\Gamma$. 

\begin{figure}[t] 
	\centering
	\includegraphics[width = 1\linewidth]{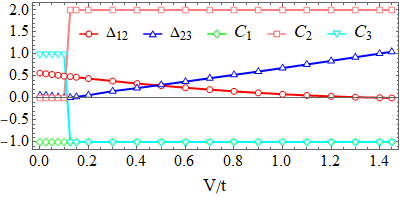}
	\caption{The Chern number of Bogoliubov mode bands in addition to the energy gap between different bands for different values of the nearest neighbor interaction $V$ is plotted. The absolute value of the energy gap between the lowest (highest) and middle energy band i.e. $\Delta_{12}=|\omega_1-\omega_2|$ ($\Delta_{23}=|\omega_2-\omega_3|$) is shown by red circle (blue up triangle) line marker. The corresponding Chern numbers from lowest to highest energy band are specified as $C_1,C_2,C_3$ with green diamond, pink square, and cyan down triangle line markers, respectively. Here we supposed $U/t=3$ and $\rho=1$.}
	\label{Fig_gap-chern}
\end{figure}

Interestingly, as shown in Fig.~\ref{Fig_gap-chern}, the condensate in $\Gamma$ is not only topologically non-trivial but also changes its topology upon tuning $V$. At very small $V$, the central band is topologically trivial, while bands 1 and 3 have Chern numbers -1 and 1. At $V/t\approx 0.125$, the gap $\Delta_{23}$ between the second and third bands closes, and all bands become topologically non-trivial: band 1 and 3 acquire both Chern number -1, whereas the central band exhibits a higher Chern number of value 2. This topology persists up to $V=U/2$, where the gap closing between bands 1 and 2 yields another topological phase transition. However, as mentioned earlier, at this value also a structural change of the mean-field occurs.

\begin{figure}[t]
	\centering
	\includegraphics[width = 1\linewidth]{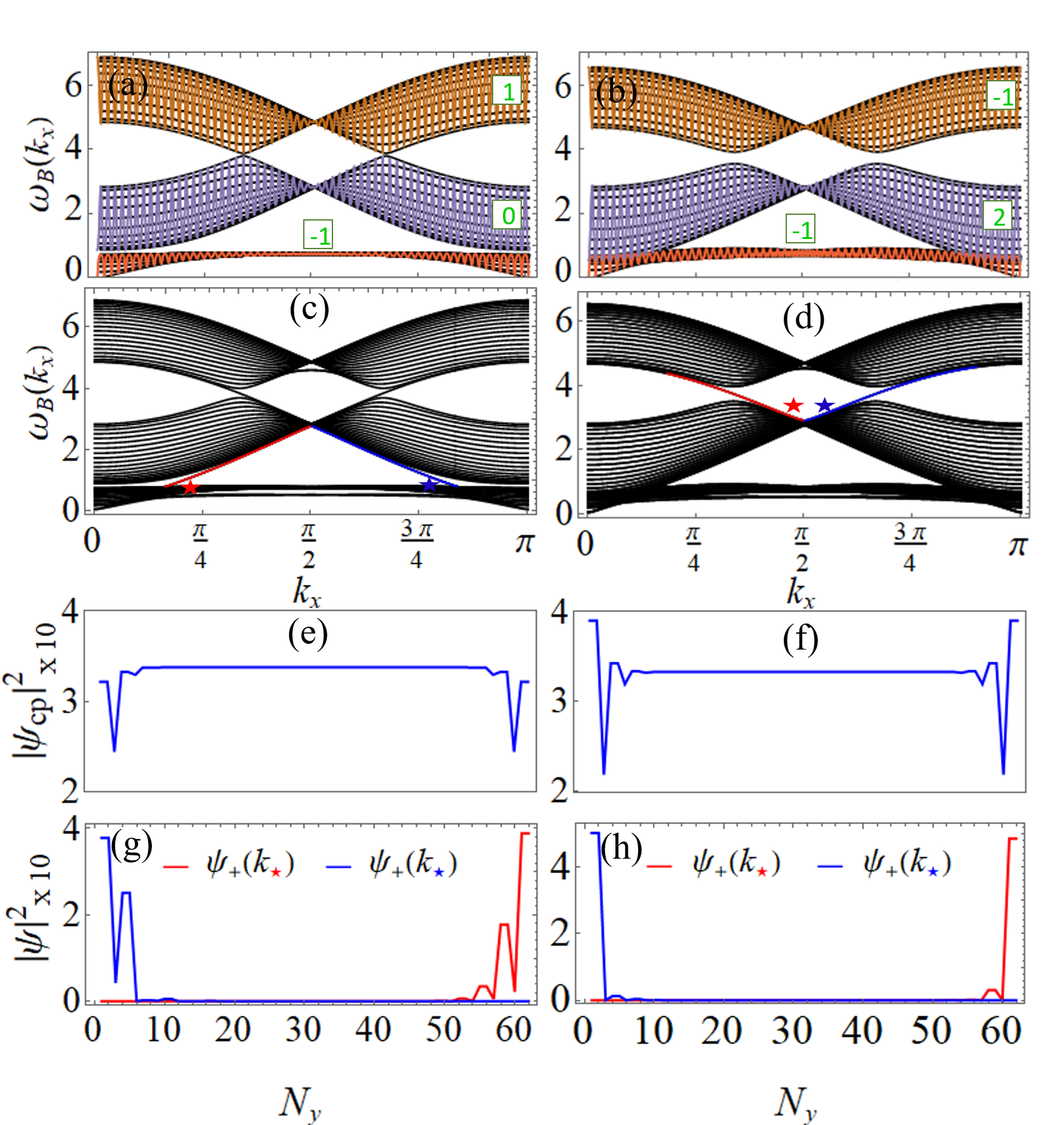}
	\caption{Dispersion relation of Bogoliubov quasiparticles above the $\Gamma$-condensate, obtained from the bulk Hamiltonian in panels (a,b), and for the slab Hamiltonian in panels (c,d). The panels on the left (a,c) are for $V=0$, and on the right are for $V/t=0.6$ while in all panels, $U/t=3$ is assumed. In the bulk Hamiltonian, all bands (in different colors) are energetically separated from each other, and Chern numbers are well defined (given in the boxes), although between the first and second band in (b) the indirect band gap becomes zero. The colored edge states appearing in the slab structure are further analyzed in panels (g,h) with respect to their density profile. We see that the red (blue) states are localized  on the upper (lower) side of the slab. Noting the opposite group velocities of red and blue edge states, seen from the dispersion in (c) or (d), we interpret the pair as one chiral mode. Note that the chirality of the mode analyzed in the left panel is opposite to the chirality of the mode analyzed in the right panel. The chirality change is due to the topological transition at finite $V$, rendering the Chern number of the central band to $2$
    in the right panels.
    In panels (e,f), the density profile of the condensates is plotted.
 }
	\label{fig_spectra}
\end{figure}

\emph{Bulk-boundary Correspondence.}
The non-trivial topology of the Bogliubov bands manifests itself also through chiral edge states. To study them we apply a slab structure bounded along the $y$ axis and composed of $A, B$ sublattices at the two ends. We exploit translational invariance along the $x$ axis and consider $N_y=62$ sites along $y$. Details of the calculation for the bounded slab structure are provided in Supplemental Material~\ref{AppB}. Our main results are shown in Fig.~\ref{fig_spectra}, for $V=0$ in the panels on the left, and for $V/t=0.6$ in the panels on the right. Within the gaps of the bulk Hamiltonian [plotted in Fig.~\ref{fig_spectra}(a+b)], the slab structure exhibits one in-gap mode [see spectra shown in Fig.~\ref{fig_spectra}(c+d)], which by their wave functions [plotted in Fig.~\ref{fig_spectra}(g+h)] as well as by the condensate profile in the slab geometry [plotted in Fig.~\ref{fig_spectra}(e+f)] can be identified as localized edge states. The pair of blue and red states, at opposite edges and with opposite group velocities, can be interpreted as one chiral edge mode. 
In the figure, we have restricted our illustration to the edge states in the larger gaps, which exhibit sharper localization properties, but we note that both gaps in Fig.~\ref{fig_spectra}(c) exhibit an edge mode with the same chirality. This agrees with the trivial Chern number of the central band at $V=0$, which from the bulk-boundary correspondence is not expected to produce any change in the edge states. On the other hand, in Fig.~\ref{fig_spectra}(d) the chirality of the mode in the second gap is opposite to the chirality of the mode in the first gap. The chirality change is expected due to the Chern number of the central band now being 2. Thus, the topology of the Bogoliubov quasiparticles is reflected by the chirality of the edge state, as expected from the bulk-boundary correspondence principle~\cite{Hatsugai_BBC,RMP_BBC,Ueda_2015}.
 
\emph{Discussion.}
We have shown that nontrivial topological Bogoliubov excitation modes occur from the $\Gamma$-point condensate due to broken TRS in the corresponding BdG Hamiltonian, but not from the $K$-point condensate. This is rather curious since condensation in the $\Gamma$-point is quite usual. Hence, the question arises whether the flatness of the band is required.  We address this question by moving the flat band from the lowest to the highest energy band by means of substituting $t\to -t$ in the non-interacting tight-binding Hamiltonian Eq.~\eqref{TB_K.eqn}. The lowest energy band then has a minimum energy at $\Gamma$. For $U\geq V$, we find uniform condensation ($d^\circ_\Gamma=(1, 1, 1)/\sqrt{3}$), whereas nonuniform condensation appears for $V>U$. Independent of the interaction parameters, both PHS and TRS is preserved in the BdG Hamiltonian, and the Bogoliubov excitations are topologically trivial. We conclude that the flatness of the lowest band is crucial to obtain the topological condensate.

However, as already found in Ref.~\cite{Zhai_PRL2012}, in the case of a flat lowest band, quantum fluctuations favor
condensation in the $K$-point which lacks the interesting topological behavior. How then can we have a system with a stable topologically non-trivial $\Gamma$-point condensate? Fortunately, there is a relatively simple mechanism which can act in favor of the $\Gamma$-point condensate: It has been shown in Ref.~\cite{Moessner2018}
that breaking the hopping symmetry in the Kagome lattice, between sites corresponding to up and down triangles, and by including mass terms, it is possible to have a substantially flat band with a controllable gap closing in the non-interacting Hamiltonian by manipulating the hopping parameters. Through this procedure, the number of condensation points can be reduced to one, at different high symmetry points ($\Gamma, K, M$), depending on the choice of  parameters. This then allows us to obtain a $\Gamma$-point condensate from the lowest flat band which is robust against quantum fluctuations, and which has the same topologically non-trivial behavior reported above, see Supplemental Material~\ref{AppC}. Moreover, the construction allows us to remove the band-touching point and separate the flat band from the other bands on the single-particle level.
Interestingly, our numerical calculation reveals that, when the non-interacting flat band is isolated from the rest of the bands, the Bogoliubov bands become topologically trivial. 

\emph{Experimental Possibilities.}
Our proposal here is experimentally feasible to realize with ultracold bosonic Dysprosium atoms (or Rydberg atoms) in an optical lattice. The contact interaction for such an atomic species is already present. Importantly, they also possess a large magnetic dipole moment of $\sim  10$ $\mu$B which makes such
atoms interact through dipole-dipole repulsion. So, tailoring a non-interacting lattice with shorter periodicity is necessary to have a generous value of nearest neighbor interactions. In fact, tuning into different topological phases of the excitations becomes possible upon varying the lattice periodicity (or the Rydberg blockade radius) in such a system. The underlying non-interacting lattice itself can be generated by overlaying two commensurate triangular optical lattices with different wavelengths, as realized in Ref.\cite{jo12}. The last ingredient necessary is a positive value of tunneling which can originate from a synthetic gauge flux as can be realized via circular lattice shaking \cite{Jotzu2014}. The measurements of the topological edge states of the exotic excitation spectrum can then be carried out using two-photon-stimulated Raman
transitions \cite{andersen06,stanescu09} which can load a macroscopic number of bosons from the condensate
directly into the topological edge states. A 
time-of-flight measurement would then confirm the presence of vortices corresponding to the chiral topological modes. 

\begin{acknowledgments} We acknowledge support from: ERC AdG NOQIA; Ministerio de Ciencia y Innovation Agencia Estatal de Investigaciones (PGC2018-097027-B-I00/10.13039/501100011033,  CEX2019-000910-S/10.13039/501100011033, Plan National FIDEUA PID2019-106901GB-I00, FPI, QUANTERA MAQS PCI2019-111828-2, QUANTERA DYNAMITE PCI2022-132919,  Proyectos de I+D+I “Retos Colaboración” QUSPIN RTC2019-007196-7); MICIIN with funding from European Union NextGenerationEU(PRTR-C17.I1) and by Generalitat de Catalunya;  Fundació Cellex; Fundació Mir-Puig; Generalitat de Catalunya (European Social Fund FEDER and CERCA program, AGAUR Grant No. 2017 SGR 134, QuantumCAT \ U16-011424, co-funded by ERDF Operational Program of Catalonia 2014-2020); Barcelona Supercomputing Center MareNostrum (FI-2022-1-0042); EU Horizon 2020 FET-OPEN OPTOlogic (Grant No 899794); EU Horizon Europe Program (Grant Agreement 101080086 — NeQST), National Science Centre, Poland (Symfonia Grant No. 2016/20/W/ST4/00314); ICFO Internal “QuantumGaudi” project; European Union’s Horizon 2020 research and innovation program under the Marie-Skłodowska-Curie grant agreement No 101029393 (STREDCH) and No 847648  (“La Caixa” Junior Leaders fellowships ID100010434: LCF/BQ/PI19/11690013, LCF/BQ/PI20/11760031,  LCF/BQ/PR20/11770012, LCF/BQ/PR21/11840013). Views and opinions expressed in this work are, however, those of the author(s) only and do not necessarily reflect those of the European Union, European Climate, Infrastructure and Environment Executive Agency (CINEA), nor any other granting authority.  Neither the European Union nor any granting authority can be held responsible for them.
\end{acknowledgments}

\bibliography{ref}

\appendix
\begin{widetext}
\section{Mean-field approximation \label{AppA}}
In this section, we present a detailed mathematical description of the quadratic BdG Hamiltonian derivation defined in Eq.~\eqref{MFH.eqn} using mean-filed approximation. We start our analysis by considering $H_I$~(i.e. interaction Hamiltonian) that includes on-site and nearest-neighbor interactions in momentum space,
\begin{equation}
    H_I=\frac{U}{2} \sum_{\substack{k,k'\\ q,\alpha}}  d^{\dagger}_{k,\alpha}d_{k-q,\alpha} d^{\dagger}_{k',\alpha}d_{k'+q,\alpha}+
    \frac{V}{2} \sum_{\substack{k,k',q,\\ \alpha,\beta,\delta}}  e^{i\textbf{q}.\boldsymbol{\delta}} d^{\dagger}_{k,\alpha}d_{k-q,\alpha} d^{\dagger}_{k',\beta}d_{k'+q,\beta},
\end{equation}
 exploit the mean-field approximation defined as $d_k=d_{k_{cp}}^\circ+\delta d_k$ to replace the definition of each operator with its mean-filed value and its fluctuation and use Hartree-Fock-Bogoliubov approximation to achieve
\begin{align}
    \label{AppMF1.eqn}
     H^{MF}_I&=  \frac{U}{2} \sum_{\substack{k,\alpha}}  4d^{\circ*}_{k_{cp},\alpha}d^\circ_{k_{cp},\alpha} \delta d^{\dagger}_{k,\alpha}\delta d_{k,\alpha}
     +d^{\circ *}_{c,\alpha}d^{\circ *}_{c,\alpha}\delta d_{k_{cp}-k,\alpha}\delta d_{k_{cp}+k,\alpha}
    +d^\circ_{k_{cp},\alpha}d^\circ_{k_{cp},\alpha} \delta d^{\dagger}_{k_{cp}+k,\alpha}\delta d^{\dagger}_{k_{cp}-k,\alpha}
     -3|d^\circ_{k_{cp},\alpha}|^2\nonumber \\
    &+\frac{V}{2} \sum_{\substack{k,q,\\ \alpha,\beta,\delta}} 
    e^{i\textbf{q}.\boldsymbol{\delta}} \left[ 2 d^{\circ*}_{k_{cp},\alpha}d^\circ_{k_{cp},\alpha} \delta d^{\dagger}_{k,\beta}\delta d_{k,\beta} \delta_{q,0}
     +d^\circ_{k_{cp},\alpha}d^\circ_{k_{cp},\beta} \delta d^{\dagger}_{k_{cp}+k,\alpha}\delta d^{\dagger}_{k_{cp}-k,\beta}
     +d^{\circ*}_{k_{cp},\alpha}d^{\circ*}_{k_{cp},\beta} \delta d^{\dagger}_{k_{cp}-k,\alpha}\delta d^{\dagger}_{k_{cp}+k,\beta}
     \right. \nonumber \\
  &\left. \quad
    +d^{\circ*}_{k_{cp},\alpha}d^{\circ}_{k_{cp},\beta} \delta d^{\dagger}_{k_{cp}-k,\beta}\delta d_{k_{cp}-k,\alpha}  +d^{\circ*}_{k_{cp},\alpha}d^{\circ}_{k_{cp},\beta} \delta d^{\dagger}_{k_{cp}+k,\beta}\delta d_{k_{cp}+k,\alpha}-3 |d^{\circ}_{k_{cp},\alpha}|^2  |d^{\circ}_{k_{cp},\beta}|^2  \delta_{q,0}  \right ].
\end{align}
We then include the tight-binding Hamiltonian to this Hamiltonian to achieve BdG Hamiltonian in Nambu space as
\begin{align}
    \label{BG-S.eqn}
   &H_B(k)=\frac{1}{2}\sum_{\alpha,\beta}\Psi^\dagger_{k,\alpha} H^{MF}_{\alpha,\beta} \Psi_{k,\beta}+\text{const.},
   & H^{MF}=
   \begin{bmatrix}
        H_0(k_{cp}+k)+\mathcal{H}_{0}(k) & H_{\Delta }(k) \\
       H^*_{\Delta }(k)  & H^T_0(k_{cp}-k)+\mathcal{H}^{ *}_{0}(k)
    \end{bmatrix},
\end{align}
in which the wave function ${\bf \Psi}_k = (\Psi_{k_+},\Psi^\dagger_{k_-})^T$, and $\Psi_{k_\pm} = (\delta d_{k_\pm,A},\delta d_{k_\pm,B},\delta d_{k_\pm,C})^T$, $k_\pm=k_{cp}\pm k$ are momentum corresponding to the particle-like $k_{+}$ and hole-like $k_{-}$. The Bogoliubov Hamiltonian in Eq.~\eqref{BG-S.eqn} also consists of the block diagonal matrix $H_0$ which generally is a tight-binding Hamiltonian in momentum space Eq.~\eqref{TB_K.eqn} and $ \mathcal{H}_{0}$ is defined as
\begin{equation}
    \label{H0MF.eqn}
    \mathcal{H}_{0}=2V 
    \begin{bmatrix}
      \left(\rho-n_A+U  V^{-1} n_A\right)&  \zeta_{AB} \cos k_1 & \zeta_{Ac} \cos k_2  \\ 
       \zeta^*_{AB} \cos k_1 &  \left(\rho-n_B+U V^{-1} n_B\right) & \zeta_{BC} \cos k_3 \\
       \zeta^*_{Ac} \cos k_2  & \zeta^*_{BC} \cos k_3 & \left( \rho-n_C+U V^{-1} n_C\right)
    \end{bmatrix},
\end{equation}
and block off-diagonal matrix can be expressed as,
 \begin{equation}
    \label{HpMF.eqn}
     H_{\Delta}=2V
    \begin{bmatrix}
       U \Delta_{AA}/2V &  \Delta_{AB} \cos k_1 & \Delta_{Ac} \cos k_2  \\ 
       \Delta_{AB} \cos k_1 &  U \Delta_{BB}/2V & \Delta_{BC} \cos k_3 \\
       \Delta_{Ac} \cos k_2  & \Delta_{BC} \cos k_3 & U \Delta_{CC}/2V
    \end{bmatrix}.
 \end{equation}
Here $\zeta_{\alpha\beta}=d^\circ_{k_{cp},\alpha} d^{\circ *}_{k_{cp},\beta}$, and $\Delta_{\alpha\beta}=d^{\circ }_{k_{cp},\alpha} d^{\circ }_{k_{cp},\beta}$ and  $\rho=n_A+n_B+n_C$.

\section{Slab structure}
\label{AppB}
In this section, we consider the slab structure of Kagome lattice that is bounded along the $y$ axis to investigate the bulk-boundary correspondence. We start our model by defining the annihilation operator as
\begin{equation}
    d_{x,y,\alpha}=\frac{1}{\tilde N_{uc}} \sum_{k_x} e^{i k_x x}  d_{y,k_x,\alpha},
\end{equation}
for $\tilde{N}_{uc}$ the number of unit cells along the $x$ axis, and then we express the slab Hamiltonian as
 \begin{align}
   \tilde H_{T}(k_x)=&\sum_{\substack{k_x,\alpha,\beta\\ <y,y'>}}d^{\dagger}_{y,k_x,\beta}H_{0,\alpha,\beta}(k_x,y,y')  d_{y',k_x,\beta}+
    \frac{U}{2}\sum_{\substack{k_x,k'_x,q,\\\alpha,y}}d^{\dagger}_{y,k_x,\alpha}d_{y,k_x-q_x,\alpha} d^{\dagger}_{y,k'_x,\alpha}d_{y,k'_x+q_x,\alpha}+\nonumber\\
    &\frac{V}{2}  \sum_{\substack{k_x,k'_x,q_x,\\ \alpha\neq\beta,\delta\\<y,y'>}} e^{i q_x\delta^x_{\alpha\beta}} d^{\dagger}_{y,k_x,\alpha}d_{y,k_x-q_x,\alpha} d^{\dagger}_{y',k'_x,\beta}d_{y',k'_x+q_x,\beta}.
    \label{tildeHBH.eqn} 
\end{align}
Here we define $\delta^x_{\alpha\beta}$ as the distance between two nearest neighbors along the $x$ axis with different sublattice indexes at positions $y, y'$ along the $y$ axis. To achieve the condensation parameters for this slab structure we exploit the homogeneous ansatz and express each bosonic operator $d_{y,k_x,\alpha}$ in terms of its mean-field value as
\begin{equation}
    \label{edge_ope.eqn}
    <\tilde d_{y,k_x}>=\tilde d^\circ_{y,k_x}=e^{i k_x x}(\psi_{y_1,\alpha_1},\dots\psi_{y_N,\alpha_N}),\quad \sum_{i=1}^N |\psi_{y_i,\alpha_i}|^2=\tilde \rho,
\end{equation}
for $y_i$ and $\alpha_i$ represent the position, and corresponding sublattice index along the edge, respectively, and $\tilde\rho$ is the number of particles in each unit cell. Replacing \eqref{edge_ope.eqn} in the slab Hamiltonian, we evaluate the mean-field energy at momentum $k$ as
\begin{align}
    \label{MFenergy_edge.eqn}
    \tilde E_{MF}(k_x)-\mu \tilde \rho =\sum_{\substack{<y,y'>\\\alpha,\beta}}& H_{0,\alpha,\beta}(k_x,y,y') \psi^{\star}_{y,\alpha} \psi_{y',k_x,\beta} +
    \frac{U }{2}|\psi_{y,\alpha}|^2 |\psi_{y',\beta}|^2 \delta_{\alpha,\beta} \delta_{y,y'} 
    +\frac{V }{2}|\psi_{y,\alpha}|^2 |\psi_{y',\beta}|^2.
\end{align}
We note that mean-filed energy minimization with respect to each wave function component $\psi_{y_i,\alpha_i}$, gives us the corresponding Gross-Pitaevskii equation, and the condensation parameter for this system can then be readily obtained using the imaginary time evolution method~\cite{Ueda_2015}. We also achieve the edge states of BdG 
Hamiltonian for this lattice structure through using the similar steps as presented in the previous section for corresponds bulk Hamiltonian commensurate with a small difference in the new definition of each operator for our proposed slab structure
in Eq.~\eqref{edge_ope.eqn}.

\section{Broken hopping symmetry}\label{AppC}
In this section, we consider a new scheme of Kagome lattice in which the hopping amplitudes between up and down triangles are controllable and there is an additional mass term for each sublattice~\cite {Moessner2018}. The corresponding non-interacting tight-binding Hamiltonian for non-symmetric Kagome lattice can be separated into two parts namely up and down triangles
\begin{equation}
    \label{TB_BS.eqn}
    \Tilde{H}(k)=H_u(k)+H_d(k),\quad
    H_d(k)=
    \begin{pmatrix}
     |t_{A,d}|^2& 
    e^{-i k_1} t_{A,d} t^*_{B,d} & 
    t_{A,d} t^*_{C,d}  e^{-i k_2 }\\
     t_{B,d} t^*_{A,d} e^{i k_1} & |t_{B,d}|^2 & t_{B,d} t_{C,d}^*  e^{ik_3}  \\ t_{C,d} t^*_{A,d} e^{i k_2}&
    t_{C,d} t^*_{B,d} e^{-i k_3} &   |t_{C,d}|^2 
    \end{pmatrix}.
\end{equation}

Switching from the symmetric to non-symmetric hopping parameters~(through Eq.~\eqref{TB_BS.eqn}) with additional mass terms, it is possible to have band energies with common properties to the symmetric Kagome lattice, such as the lowest flat band energy along with gapless energy bands which comes up with only one condensation point degeneracy at $k_{cp}=\Gamma$ independent of the value of interaction parameters. Tuning the nearest neighbor interaction parameter for constant on-site interaction $U/t=3$ we can investigate the topological properties of Bogoliubov excitation modes presented in Fig.~\ref{Gap_Chern_moess.fig}. In this figure, we assumed two sets of hopping parameters so that all single-particle bands are gapless and the lowest band is flat.  Both panels share the same properties as the symmetric Kagome lattice in Fig.~\ref{Fig_gap-chern} and the only difference is that by means of symmetry breaking we can remove $K$-point condensation with trivial topological properties.
  \begin{figure}[t]
	\centering
	\includegraphics[width = 0.9\linewidth]{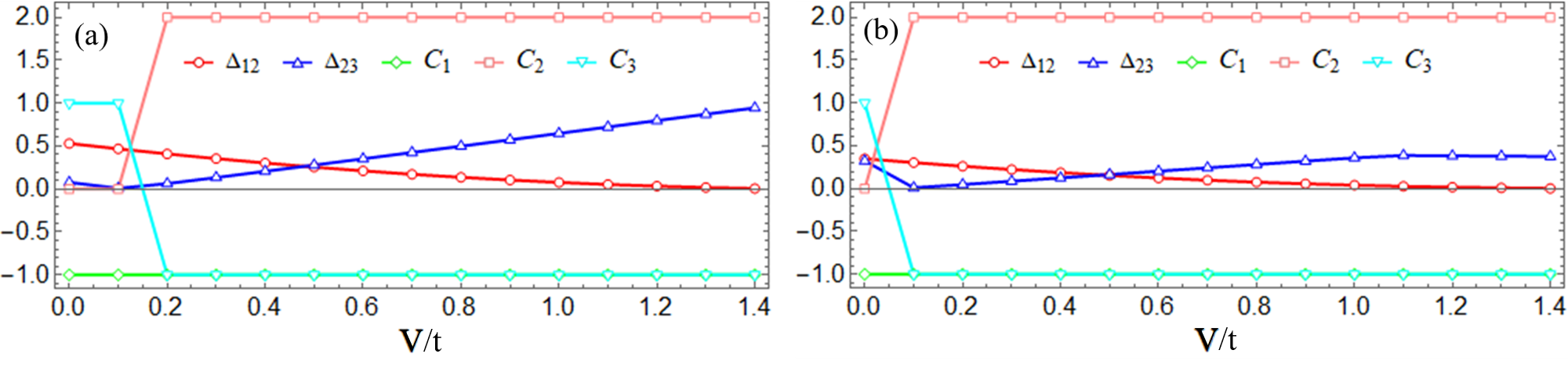}
	\caption{Chern number of Bogoliubov bands in addition to the pa properties of them for two different sets of hopping parameters have been depicted. In the left panel hopping parameters are $x=0.95t,  t_{Au}=t_{Ad}=1/x, t_{Bu}=t_{Bd}=t_{Cu}=t_{Cd}=x$. The defined parameters in right panel is $x=t, t_{Au}=t_{Ad}=t_{Bu}=t_{Bd}=x, t_{Cu}=t_{Cd}=0.5x$. The considered on-site Bose-Hubbard interaction is $U/t=3$ }
	 \label{Gap_Chern_moess.fig}
\end{figure}

\end{widetext}

\end{document}